\documentstyle[aasms4,12pt]{article}
\newcommand{\Msolar}{\mbox{\,$M_{\odot}$\/}}          
\newcommand{\HII}{\mbox{H\,{\footnotesize II}}}       
\newcommand{\magap}[1]{\mbox{$ #1^m$}}                     
\newcommand{\filter}[1]{\mbox{\it #1\/}}              
\newcommand{\oversim}[2]{\lower0.5ex\vbox{\baselineskip=0pt\lineskip=0.2ex
     \ialign{$\mathsurround=0pt #1\hfil##\hfil$\crcr#2\crcr\sim\crcr}}} 
\newcommand{\simless} {\mbox{$\mathrel{\mathpalette\oversim<}$}} 
\newcommand{\etal}{\mbox{\hbox{\it et\,al.}}}         
\newcommand{\eg}{\mbox{\hbox{\it e.g.},}}             
\newcommand{\etcetera}{\mbox{\hbox{\it etc.}}}        
\newcommand{\idest}{\mbox{\hbox{\it i.e.},}}          
\newcommand{\cf}{\mbox{\hbox{\it cf.}}}               
\newcommand{\kmpers}{\mbox{\,km\,s$^{-1}$}}           
\slugcomment{In Press,  Astronomical Journal, January 1997}
\begin{document}
\title{The OMC-1 molecular hydrogen outflow \\
       as a fragmented stellar wind bubble}
\author{Mark J. McCaughrean \& Mordecai-Mark Mac Low \\
Max-Planck-Institut f\"ur Astronomie, \\
K\"onigstuhl 17, 69117 Heidelberg, Germany \\
Electronic mail: mjm \& mordecai@mpia-hd.mpg.de
}
\begin{abstract}
We present new images of the OMC-1 molecular hydrogen outflow, made
using long-slit spectroscopy in order to accurately subtract the
underlying continuum emission. These images reveal an extremely
clumpy, quasi-spherical inner shell that breaks up at larger radii
into bow-shocks and trailing wakes in the north-west, as originally 
described by Allen \& Burton (1993); a fainter counter-finger
to the south-east is newly discovered in the present data. While 
the outflow appears to be broadly bipolar, this is probably due to 
an interaction between an initially spherical wind from the source 
and a large-scale density enhancement surrounding it, 
rather than direct collimation imposed close to the source. The clumpy 
appearance of the inner shell confirms the prediction of the recent model 
of Stone, Xu, \& Mundy (1995), in which a spherical and time-varying 
wind fragments a swept-up shell, producing high-velocity shrapnel, which 
in turn drives bow-shocks into the surrounding gas, resulting in the 
observed ``fingers''. As an alternative to the single varying source 
proposed by Stone \etal, we speculate that several young sources in 
the BN-KL cluster may have been responsible for first sweeping up the shell 
and then fragmenting it. \\
%
\end{abstract}

\section{Introduction}
Mass outflow is known to be a common and perhaps inevitable part
of star formation, a process more axiomatically associated with
mass infall. Observations of young low-mass stars at optical, near-infrared, 
and millimeter wavelengths often reveal highly collimated bipolar 
jets and molecular outflows (\cf{} Edwards, Ray, \& Mundt 1993), generally 
believed to be driven by magnetocentrifugal forces arising through an 
interaction between the magnetic field and wind associated with a 
rotating circumstellar disk (\eg{} K\"onigl \& Ruden 1993). The situation 
for high-mass stars is less clear: although winds are known to be a 
continuing force throughout the evolution of a massive star (\cf{} Langer 
\etal{} 1994), their role is not well understood at the earliest phases, 
at least in part because massive stars are intrinsically rather rare and 
generally quite far from the Sun, making it harder to study them in detail. 
Nevertheless, bipolar outflows have been found from a few young massive 
stars including Cepheus~A (Bally \& Lane 1991), S\,106 (Staude \& Els\"asser 
1993), IRAS\,08159$-$3543 (Neckel \& Staude 1995), G25.65+1.05 and
G240.31+0.07 (Shepherd \& Churchwell 1996). Compared to low-mass systems, 
their outflows tend to be more chaotic, spreading over a wider angle, and 
often appearing to have multiple jets. As yet however, the driving 
mechanisms for the outflows from young massive stars remain unclear. 

The nearest site of ongoing massive star formation is in the BN-KL 
complex embedded in the OMC-1 molecular cloud behind the Orion Nebula 
(Genzel \& Stutzki 1989). IRc2 has long been thought to be the dominant 
source of luminosity in the complex, in part because it is the brightest 
compact source of mid- to far-infrared luminosity, and also because of
its association with typical high-mass star formation 
tracers such as water masers (Wynn-Williams \etal{} 1984). However, recent 
high spatial resolution infrared and radio observations have shown the 
situation to be considerably more complicated (Gezari 1992; Dougados \etal{} 
1993; Menten \& Reid 1995). When viewed at sub-arcsec resolution, the 
nominal point source IRc2 is seen to break up into a small cluster of 
sources, none of which are directly associated with radio emission 
from ultracompact \HII{} regions expected of young, massive stars. 
However, such an ultracompact \HII{} region is found to lie directly south 
of the infrared sources, suggesting that perhaps the latter are all 
simply ``holes in the clouds'' illuminated by the true dominant
luminosity source (Menten \& Reid 1995), and that luminosity estimates 
based on their infrared emission are unreliable. Furthermore, the 
nearby source ``n'' (Lonsdale \etal{} 1982) shows both strong IR emission 
and radio emission from an ultracompact \HII{} region with a hint of 
bipolar structure that lies within the error bars for the center of 
emission for the water masers, unlike IRc2 (Menten \& Reid 1995). 
Therefore, it seems apparent that the core of the BN-KL complex holds 
a small cluster of very young self-luminous sources, also including 
the BN and IRc9 sources (Beck 1984). What remains uncertain is which 
source, if any, dominates the total luminosity of the complex.

What is certain, however, is that the BN-KL complex is associated
with a spectacular outflow. Embedded in OMC-1 and obscured by 
$A_V$\,$\sim$\,\magap{20}--\magap{50}, the outflow must be observed at 
infrared wavelengths or longer: the shocked molecular gas associated 
with the outflow is best seen in the near-infrared emission lines of 
H$_2$, notably the v=1--0 S(1) line at 2.122\micron{} (Beckwith \etal{} 
1978; Taylor \etal{} 1984). The increased spatial resolution achieved 
by infrared imaging observations over the past decade has revealed 
a striking swath of bow-shocks and trailed wakes (henceforth referred 
to collectively as ``fingers'') originating in the vicinity of IRc2 
(Lane 1989; Allen \& Burton 1993; Sugai \etal{} 1994, 1995; 
Everett, DePoy, \& Pogge 1995; Schild, Miller, \& Tennyson 1996).

The multiple fingers of shocked H$_2$ cover a wide angle on the sky and 
therefore pose a problem for most models developed for outflows from 
low-mass stars, in which only a single jet is thought to be active. 
It is conceptually possible for a single precessing jet to create the 
multiple fingers, but as the cooling time for excited H$_2$ is short 
(on the order of one year), the precession speed required to produce 
multiple fingers would be extraordinary; furthermore, curved jets would 
be expected (Schild \etal{} 1996). With these problems in mind, Allen 
\& Burton (1993) revived a long dormant model for Herbig-Haro objects, 
in which ``bullets'' of material are ejected from a central source, 
forming bow-shocks and shocked wakes behind them as they plow through 
ambient molecular gas. Allen \& Burton suggested that a single explosive 
event in the BN-KL complex within the past few thousand years might have 
resulted in a veritable barrage of ``artillery shells'' (Burton \& Allen 
1994), creating the swath of fingers seen today. However, the bullet model 
for H-H objects, bow-shocks, and wakes had fallen out of favor because of 
the difficulty associated with ejecting isolated clumps of molecular gas 
from a source in an explosive event and having them remain coherent over 
long paths through the surrounding molecular cloud. In the case of OMC-1, 
there is also the problem that multiple bullets must be ejected almost 
simultaneously over a wide fan. 

Stone, Xu, \& Mundy (1995) recently proposed a new model for creating the 
bullets and swath of fingers seen in OMC-1. Their model starts with a 
central, massive, young stellar object driving a strong spherical wind, 
sweeping up ambient gas into a thin, dense shell. As long as such a shell 
continues to decelerate, it will remain stable to Rayleigh-Taylor 
instabilities.  However, Stone \etal{} propose that the wind velocity may 
vary with time, and that if the velocity were to increase sharply, the shell 
would be accelerated into the surrounding dense medium, become subject to 
Rayleigh-Taylor instabilities, break into fragments, and form exactly 
the swath of bow-shocks and wakes that is observed. Thus, their model 
achieves three important things: first, fragments are created; second, a 
single event at the source --- the increase in wind speed --- naturally 
leads to multiple fragments over a wide area; and third, these fragments 
are created at some finite radius from the source, thus allowing them to 
be observed quite some way from it before being destroyed. Moreover, this 
model predicts that behind the fingers, the inner part of the shell should 
break up into a highly clumpy structure, a feature that would not be 
expected in a standard bullet model, and that had not been observed 
when Stone \etal{} made their proposal. In this paper, we present new 
images of the shocked H$_2$ emission in OMC-1 which confirm that just 
such a clumpy shell lies at the base of the fingers, therefore adding 
considerable weight to the proposal of Stone \etal{} 
%

\section{Observations}
All previous images of the H$_2$ emission in OMC-1 have been made 
in a direct imaging mode, placing a narrow-band filter in front of 
a single raster-scanning detector or an imaging array. The problem 
with this technique is subtraction of bright continuum emission in 
the same band-pass from the BN-KL complex and foreground Orion Nebula. 
Observations at a nearby continuum wavelength are required, but by 
necessity must be taken before or after the H$_2$ line observations. 
Slight changes in the seeing and sky background emission are inevitable, 
making accurate removal of point and extended continuum sources difficult.

We have taken another approach to the problem, namely long-slit
spectroscopy, in which the line and continuum emission at a given 
point on the sky are measured strictly simultaneously. We used 
the MAGIC 256$\times$256 pixel near-IR camera (Herbst \etal{} 1993) 
on the Calar Alto 
2.2\,m telescope on 20 November 1994 with a pixel size of 0.6 arcsec, 
and a slit size of 154$\times$1.2 arcsec. A direct-ruled ZnSe grism 
was used to cover the whole \filter{K} band (2.0--2.4\micron) at a 
resolution of 300, including several lines of molecular hydrogen, 
ionized hydrogen and helium, as well as adjacent continuum. A data 
cube (RA$\times$dec$\times$$\lambda$) was built up by stepping the 
slit over the region, one slit width at a time, with an
integration time of 60 seconds per position. Equivalent sky observations 
were made every 30 minutes several arcmin to the west. The total number 
of on-source slit positions was 120, therefore covering a total area of 
$\sim$\,2.5$\times$2.5 arcmin with an effective pixel size of 1.2$\times$0.6 
arcsec. The region is shown in Figure~1 (Plate~A) overlaid on a 
narrow-band continuum (\idest{} no line emission from H$_2$ 
\etcetera) image with a similar spatial resolution ($\sim$\,1.5 arcsec 
FHWM) for comparison; key objects are labeled for reference.

The data reduction process is illustrated in Figure~2 (Plate~B). 
Figure~2a shows a single raw source frame: the vertical lines are
from the extended nebular gas and OH airglow which dominates the sky 
background at these wavelengths; the horizontal stripes are from 
compact continuum sources (stars). Figure~2b is a single sky frame, 
showing just the OH airglow lines. Figure~2c is the result of subtracting
a stack of sky frames from the source frame. In this image, the vertical
lines are slightly curved, the horizontal lines slightly tilted, and 
the wavelength dispersion non-linear, due to various instrumental effects. 
The curvature and dispersion were determined as a function of slit 
position and wavelength by fitting bright OH airglow lines in the sky 
frames, and the continuum tilt was measured from a number of bright stellar 
spectra. The data were then rectified to give a linear wavelength 
dispersion, with extended emission lines running down columns and 
discrete continuum sources across rows. Each image was then divided 
by a tungsten-illuminated spectral dome flat. Similar data processing 
was applied to spectra of a bright spectral standard star, which were 
then extrapolated to form a full spectral image. Finally, each source 
image was divided by the standard star image, then multiplied by the
appropriate black-body curve in order to remove the atmospheric transmission 
profile. Figure~2d shows the final calibrated and rectified source
data for one slit position. Several molecular and ionized nebular 
emission lines are identified for reference, although only the v=1--0 S(1) 
H$_2$ line is discussed further in this paper.

The spectral images were stacked to form a cube with axes RA, dec, and 
wavelength, thus allowing us to extract RA$\times$dec slices at any 
given wavelength. Three slices straddling the H$_2$ v=1--0 S(1) line 
at 2.122\micron{} were extracted and co-added, weighted by intensity;
an equivalent continuum was made by averaging five slices on either side 
of 2.122\micron{}. These two images were bilinearly interpolated up by 
a factor of 4$\times$2, in order to give images with equal pixel size 
(0.3 arcsec) in RA and declination (Figures~3a and 3b; Plate C). The 
vertical stripes associated with bright point sources arise due to 
scattering in the camera, but subtract out well as seen below. The 
effective spatial resolution in these images is $\sim$\,1.5 arcsec FWHM\@.
Finally, the continuum image was subtracted from the 2.122\micron{} 
image to give a pure H$_2$ v=1--0 S(1) line image (Figure~4, Plate~D).
The locations of the bright Trapezium OB stars, BN, and IRc2 are marked 
with white circles. Intensity has been scaled logarithmically to 
enhance faint details, and residual images of some of the brighter 
stars are visible due to small interpolation errors. 

\section{Discussion}
The bow-shocks and wake fingers described by Allen \& Burton (1993) are 
seen to the north-west in the new H$_2$ image shown in Figure~4,
although they are not completely covered. As noted by Schild 
\etal{} (1996), there are also a number of 
short fingers in the south-east; our new data reveals for the first time
however, a direct counterpart to the north-west fingers in the form 
of a large faint bow-shock and wake extending over 80 arcsec (0.17\,pc) 
above the Trapezium OB stars. Inside the fingers, closer to their origin, 
the H$_2$ emission is extremely clumpy. This feature was predicted by 
Draine \& McKee (1993) on the basis of the largely invariant H$_2$ line 
ratios measured with large aperture spectrometers (Brand \etal{} 1988, 1989). 
These line ratios should vary considerably from tip to wake of a bow-shock, 
and thus for them to appear constant, Draine \& McKee (1993) argued that 
there must be many such bow-shocks within a spectrometer beam, and 
that the entire outflow must contain hundreds of clumps of diameter 
$\simless$\,1~arcsec or so. The model of Stone \etal{} (1995) naturally 
predicts such a clumpy shell inside the elongated fingers as shown in 
Figure~5. (It must be remembered that the model is only a two-dimensional 
slice through the outflow, while the observed H$_2$ image in Figure~4 
shows the full three-dimensional structure projected on the sky.) 

This comparison of model and observations brings into focus the vexing 
issue of the underlying geometry of the OMC-1 outflow. The Stone \etal{} model 
starts with an idealized spherically symmetric wind, while in contrast, 
observations clearly show the real outflow to be elongated along the 
north-west/south-east axis. Thus it has typically been described as bipolar 
(\eg{} Sugai \etal{} 1994, 1995; Erickson \etal{} 1982; Schulz \etal{} 1995). 
However, this apparent incompatibility between the data and model can be 
reconciled by noting that the OMC-1 outflow is probably not 
{\em intrinsically\/} bipolar: it is unlikely that direct collimation 
occurs within a few AU of the source as is thought to be the case in 
low-mass outflows. The region around the nominal IRc2 includes a dense 
ridge of quiescent gas and dust with a hot core which runs from north-east 
to south-west, \idest{} perpendicular to the dominant high-velocity 
(hundreds of \kmpers) outflow axis (Murata \etal{} 1992; Genzel \& 
Stutzki 1989). This ridge is associated with a low-velocity (tens of
\kmpers) outflow seen in molecular lines including SO, SiO, HCO$^+$, 
and CS (Plambeck \etal{} 1982; Wright \etal{} 1983; Vogel \etal{} 1984; 
Murata \etal{} 1991), in which clumps of gas are thought to be in 
the form of an ``expanding doughnut'' around IRc2 with a radius 
on the order of thousands of AU (Plambeck \etal{} 1982). It is likely
that the dense ridge and ``doughnut'' are some remnant of the original 
molecular cloud structure, with some suggestion that they may also 
be related to a putative circumstellar disk or torus around the outflow 
driving source (Murata \etal{} 1991). It seems as though the initially 
spherical shell and outflow interact with this 
large-scale density enhancement, and are then shaped into a more elliptical 
form, with perhaps some pinching at the waist (Sugai \etal{} 1994, 1995). 
Further out, Sugai \etal{} (1995) suggest that the fingers arise as ``some 
parts of the outflow penetrate the expanding shell probably due to some 
inhomogeneities in the outflow and/or in the shell'': it now seems 
clear that Rayleigh-Taylor instabilities are responsible for the 
transition from a shell to the fingers, as modeled 
by Stone \etal{} and discussed below in Section~4.


The degree of shaping or effective collimation in the flow is, at present, 
uncertain, as it depends on the angle of flow with respect to the 
line-of-sight. Most observational results suggest that the molecular 
outflow is tilted out of the plane of the sky, with the north-west 
``lobe'' pointing close to the line-of-sight towards the observer 
(Schulz \etal{} 1995). This appears to be confirmed by the detection 
at optical wavelengths of [OIII] emission associated with the tips of 
the north-west fingers, which are delineated at IR wavelengths by high 
excitation [FeII] emission (Axon \& Taylor 1984, Allen \& Burton 1993). 
To be detected in the optical, these tips must be emerging from the OMC-1 
molecular cloud into the foreground Orion Nebula, and indeed, the [OIII] 
emission is highly blue-shifted (Axon \& Taylor 1984). By contrast, the 
receding south-east ``lobe'' may be pushing into denser material, 
accounting for its shorter and fainter appearance. However, based on 
their mid-IR H$_2$ line profiles, Parmar \etal{} (1994) suggest the 
underlying outflow is very nearly {\em in\/} the plane of the sky, so 
the orientation remains somewhat ambiguous. However, it seems as though 
the degree of effective collimation in the outflow must be relatively
small regardless, \idest{} that the outflow has a wide opening angle 
(\cf{} Geballe \etal{} 1986). A highly collimated flow almost along 
the line-of-sight would be strongly fore-shortened and show few if any 
long fingers, in contrast to the wide swath of long H$_2$ fingers 
actually seen. On the other hand, a flow in the plane of the sky 
should not show high velocity blue shifted emission unless it was 
again only loosely collimated. Future measurement of proper motions 
and radial velocities for the H$_2$ fingers will make it possible to 
determine the degree of collimation and the geometry of the flow with 
respect to the plane of the sky.

\section{Analysis}
Stone \etal{} (1995) found that the bullets produced in their numerical 
model decelerated rapidly, and therefore they had to invoke a density 
gradient in the surrounding medium to explain the observed velocities. 
However, it appears that this rapid deceleration was a consequence of 
their adopting the extremely high sound speed of 20\kmpers{} in their 
ambient gas in order to maintain numerical resolution in the shell 
modeled by their computations. Next, we discuss the consequences of 
setting the sound speed to a more reasonable value.
 
The characteristic time for stopping a dense cloud or bullet is
of order $10 t_c$, where the cloud-crushing time $t_c$ is the time 
required for a shock to propagate entirely through the cloud, and 
is given by
\begin{equation}
t_c = \chi^{1/2} r_b / v_b,
\end{equation}
where $\chi = \rho_b/\rho_0$ is the ratio of the bullet density to that 
of the ambient gas, $r_b$ is the radius of the bullet, and $v_b$ is its 
velocity through the background gas (Klein, McKee, \& Colella 1994). For 
an isothermal shock, the compression ratio $\chi$ is given by the square 
of the Mach number, so $\chi = {v_{b_{0}}}^2/{c_s}^2$, where $v_{b_{0}}$ 
is the initial velocity of the bullets after fragmentation of the shell 
and $c_s$ is the sound speed in the gas. As a strong lower-limit on 
$v_{b_{0}}$, we can take the present-day greatest bullet velocity of 
$v_b \sim 400$\kmpers{} from Allen \& Burton (1993), from whom we can 
also take the mass $m_b$ of a typical bullet as $10^{-6}$\Msolar. The 
excitation temperature of the CO in the shell lies between 60--90\,K 
(Snell \etal{} 1984), corresponding to sound speed $c_s \sim 0.5$\kmpers.
Magnetic fields could raise the effective sound speed to as much as 
several \kmpers, but no more.
 
We assume that the mass $M$ of the gas responsible for the present day
high-velocity CO emission is a rough indication of the mass that was
present in the spherical shell just before it fragmented. Also, we
assume a radius $R$ of $\sim$\,20 arcsec as an estimate of the present
shell size based on the transition from the inner clumpy shell to
fingers in the H$_2$ image.  From these numbers, we can crudely
estimate the local ambient density $\rho_0$ when the shell fragmented
as $\rho_0 = 3 M / 4\pi R^3$.  Taking $M = 8.2$\Msolar{} (Snell
\etal{} 1984) and a distance to the BN-KL complex of 480\,pc (Genzel
\etal{} 1981), we find $\rho_0 \sim 1.6 \times 10^{-18}$\,g\,cm$^{-3}$.

Finally, we may rewrite the equation for the cloud-crushing time in terms 
of the typical bullet velocity and mass, shell density, and sound speed, 
to find
\begin{eqnarray}
t_c &   =   & (3/4\pi)^{1/3} v_b^{2/3} c_s^{-5/3} \rho_0^{-1/3} m_b^{1/3} \\
    & \sim  & (1.4 \times 10^4 \mbox{ yr}) v_{400}^{2/3} c_{km}^{-5/3}
              \rho_{-18}^{-1/3} m_{-6}^{1/3},
\end{eqnarray}
where 
$v_{400} = v_b$/(400\kmpers), 
$c_{km}$ is the sound speed in \kmpers,
$\rho_{-18} = \rho_0$/(10$^{-18}$\,g\,cm$^{-3}$), and 
$m_{-6} = m_b$/(10$^{-6}$\Msolar).  
As Stone \etal{} found, sound speeds of 20\kmpers{} will result in rapid 
deceleration due to the low density of the resulting bullets. However, 
realistic effective sound speeds of only a few \kmpers{} will 
result in bullets dense enough to propagate much farther than is 
presently observed without strong deceleration. Therefore, while we 
do not rule out a density gradient in the molecular cloud core 
surrounding the outflow (indeed, one is suggested by the shape of the 
outflow and by the observation of optical emission from the outermost 
bullets), we find that no gradient is {\em required\/} to explain 
the bullets and fingers.

It is worth considering the issue of how the H$_2$ emission actually 
arises in the fingers. Schild \etal{} (1996) compute the thickness of 
the observed layer of warm (T\,$\sim$2000\,K) H$_2$ in the structure 
they call the ``north jet'', finding it to be around 0.05\% of the 
finger cross section. They speculate that this thin layer is formed 
as ambient H$_2$ is entrained by some unseen jet of material, then 
heated above its dissociation temperature. However, within the Stone 
\etal{} (1995) model, there {\em is\/} no jet, and the natural 
explanation is that behind the high-velocity fragment, an expanding 
wake shocks ambient H$_2$, which then rapidly cools below observable 
temperatures (Hollenbach \& McKee 1989). Schild \etal{} also note that 
water masers have been observed coincident with fingers and with large 
proper motions pointing outwards in the same direction as the fingers. 
Water masers are most easily produced by shocks in dense gas (Elitzur, 
Hollenbach, \& McKee 1989), so this observation is also naturally 
explained by the fingers being bounded by shocks, as Stone \etal{} predict.

Therefore, on balance, we find that with suitable minor modifications,
the Stone \etal{} model does a good job of describing how the OMC-1 
H$_2$ outflow is formed, with our new images providing further 
observational support. There is, however, one {\it ad hoc\/}
component in the Stone \etal{} model, namely a spherical wind from 
the driving source that is strongly time variable. There is little 
observational evidence for strong spherical winds from massive YSOs 
on the large scales required (although several massive YSOs are known 
to have strong winds much closer in; Hoare \etal{} 1994). It follows 
that there is also little evidence for the strong time variability 
required by Stone \etal{} to accelerate the shell and reverse the 
effective gravity, leading to Rayleigh-Taylor instabilities and the 
resulting bullets and fingers. 

Schild \etal{} (1996) suggest that the need for time variability could 
be avoided if the fingers were produced by the thin-layer instability 
described by Dgani, Walder, \& Nussbaumer (1993). However, as pointed 
out by Vishniac (1994), this instability depends on the existence of a 
bow-shock-like global geometry, and so cannot explain the presence of 
instabilities at all points on the inner quasi-spherical bubble. The 
nonlinear thin shell instability described by Vishniac (1994) is 
potentially relevant as it can fragment even a decelerating 
shell if the inner, driving shock is radiative (Garc\'{\i}a-Segura, 
Mac Low, \& Langer 1995), as would be likely at the densities expected 
within a molecular cloud. However, no scenario has yet been proposed 
and simulated that can produce the observed high-velocity bullets and 
lower-velocity clumpy shell solely with the nonlinear thin shell instability.

Returning to the basic model of Stone \etal{} (1995), we propose 
a modification that introduces time-variability in an outflow not through
variations in a single source, but through non-synchronous formation
and evolution of a cluster of sources. As noted in Section~1, recent 
observations have shown that the BN-KL complex appears to contain a 
small cluster of very young high-mass stars in the vicinity of the 
nominal IRc2, plausibly an extremely youthful analog of the nearby
Trapezium OB sub-group which illuminates the Orion Nebula. Thus, our 
suggestion is that perhaps the shell or bubble described by Stone 
\etal{} was not created by a single stellar wind, but rather by two 
or more sources in the BN-KL complex at different stages of evolution. 

In this scenario, a relatively weak source or group of sources may have 
swept up the original shell. The sharp increase in wind strength required 
to drive the Rayleigh-Taylor instabilities might then arise as a single 
new source (for example, the deeply embedded source illuminating IRc2) 
collapses, becomes luminous, and drives a powerful new wind into the 
bubble. The dynamical timescale required to create the present-day 
H$_2$ fingers is short, on the order of only $\sim$\,10$^3$\,yrs (0.2\,pc
at 200\kmpers). Therefore, since the deeply embedded young stars may be 
up to 10$^4$\,yrs old, the necessary ``desynchronization'' is relatively 
small, and the cluster could still be considered quasi-coeval. 

\section{Conclusions}
Our new continuum-subtracted H$_2$ line image of the OMC-1 outflow confirms 
the clumpy inner shell structure predicted by the time-variable wind 
model of Stone \etal{} (1995), and in addition, shows a new bow-shock 
and wake system to the south-east. The outflow appears to begin with 
a quasi-spherical wind as in the model of Stone \etal{}, but is 
subsequently loosely shaped or collimated by a large-scale density 
enhancement around the entire cluster of wind sources (Sugai \etal{} 
1994, 1995). While the degree of effective collimation and the flow 
geometry remain uncertain at this time, high-resolution near-infrared 
imaging in H$_2$ line using NICMOS on the Hubble Space Telescope and/or 
ground-based adaptive optics should allow us to measure proper motions 
of the knots, and in combination with radial velocities measured from 
spectroscopy, enable us to create a three-dimensional picture of the 
outflow. Furthermore, such high-resolution images should allow us to 
characterize the clumpy structure of the inner shell on scales a full 
order of magnitude smaller than the present data. The ultimate goal 
would be to create a ``time history'' for the outflow, in order to 
determine which source or sources were responsible for the initial 
shell and its subsequent fragmentation. In addition, sensitive imaging 
at mid-infrared wavelengths (10--20\micron{} and beyond) at high 
spatial resolution on the new generation of 8-m class telescopes may 
help better reveal the geometry and energetics of the cluster of 
potential driving sources themselves. 

\acknowledgements
We would like to thank Tom Megeath for donating the 2.2\,m time used 
to obtain the data presented here, Jianjun Xu and Jim Stone for providing 
Figure~5, and Michael Burton, C.~Robert O'Dell, Michael Smith, Alain 
Lioure, Chris Davis, Antonio Chrysostomou, and Karl Menten for 
interesting and useful discussions on the nature of the H$_2$ 
outflow in OMC-1 and its driving source.

\newpage

\centerline{FIGURE CAPTIONS}

\figcaption{%
Narrow-band 2\micron{} continuum image covering $\sim$\,3.0$\times$3.5
arcmin of the Orion Trapezium and BN-KL complex. No bright line 
emission is included in the filter bandpass. The resolution is 
1.5 arcsec FWHM\@. The region covered by the slit-scan spectroscopy 
is outlined, and the locations of key point sources are labeled 
(IRc2 is not detected at 2\micron).
}

\figcaption{%
Illustration of the slit-scan data reduction process. Fig.\ 2a shows 
a single raw spectral image taken in the Orion Nebula. Continuum point 
sources (stars) run across the wavelength dispersion or x-axis, while 
extended emission line sources (including the sky background OH airglow 
emission) run down the slit or y-axis. The slit length is 154 arcsec, 
and the wavelength range covered is 2.0--2.4\micron{} at R=300. 
Fig.\ 2b shows an equivalent raw spectral image of blank sky: only 
the OH airglow emission lines are now seen. Fig.\ 2c shows the simple
subtraction of the source and sky images, leaving just the stars across
the x-axis and and nebula emission down the y-axis. Fig.\ 2d shows
the same data after it has been rectified for curvature, tilt, and 
non-linear wavelength dispersion; flat-fielded; and corrected for
atmospheric absorption. Key nebular emission lines are labeled for 
reference. 
}

\figcaption{%
Image slices from the RA$\times$dec$\times$$\lambda$ datacube covering 
the region outlined in Fig.\ 1. Fig.\ 3a shows the average of the three
wavelength slices straddling 2.122\micron, including the H$_2$ v=1--0
S(1) line and underlying continuum emission. Fig.\ 3b shows an
average of five continuum slices either side of 2.122\micron.
Both images have been interpolated by a factor of 4$\times$2
to give a uniform pixel size of 0.3$\times$0.3 arcsec. The extended 
vertical features are artifacts due to scattering of the brightest 
continuum sources (\eg{} the Trapezium OB stars and the BN object) off 
the slit. 
}

\figcaption{%
The pure H$_2$ v=1--0 S(1) 2.122\micron{} line emission image
(continuum subtracted). The intensity has been logarithmically
stretched to show the full dynamic range. The clumpy inner
shell surrounding IRc2 is clearly seen, as are the bright
fingers of Allen \& Burton (1993) extending to the north-west.
A fainter corresponding bow-shock and wake is seen to the
south-east, passing above the Trapezium. Locations of important
continuum point sources are marked.
}

\figcaption{%
The fragmentation model of Stone \etal{} (1995). 
The image is the logarithm of the density distribution in their
two-dimensional model, showing the fragmented shell resulting from a
time-variable stellar wind. Low densities are black and high densities 
are white. The model was computed using an Eulerian hydrocode implementing 
the piecewise parabolic method. Note that this model is a two-dimensional 
cut through the shell, and cannot be compared directly to the data shown
in Figure~4, which is a two-dimensional {\em projection\/} of the 
three-dimensional OMC-1 outflow. The model bullets driving the 
outermost bow-shocks would be dense enough to travel indefinitely if 
they cooled to realistic temperatures, as discussed in the text.
}


\begin{thebibliography}{}
%
\newcommand{\nature}{{\it Nature\/}}
%
\bibitem[Allen \& Burton 1993]{allen93}
Allen, D.~A. \& Burton, M.~G. 1993, \nature, {\bf 363}, 54
%
\bibitem[Axon \& Taylor 1984]{axon84}
Axon, D.~A. \& Taylor, K.~N.~R. 1984, \mnras, {\bf 207}, 241
%
\bibitem[Bally \& Lane 1991]{bally91}
Bally, J. \& Lane, A.~P. 1991, in {\it Astrophysics with infrared arrays\/},
ASP Conf.\ Ser.\ {\bf 14}, 273
%
%
\bibitem[Beck 1984]{beck84}
Beck, S. 1984, \apj, {\bf 281}, 205
%
\bibitem[Beckwith \etal{} 1978]{beckwith78}
Beckwith, S.~V.~W., Persson, S.~E., Neugebauer, G., \&
   Becklin, E.~E. 1978, \apj, {\bf 223}, 464
%
\bibitem[Brand \etal{} 1988]{brand88}
Brand, P~W.~J.~L., Moorhouse, A., Burton, M.~G., Geballe, T.~R.,
   Bird, M., \& Wade, R. 1988, \apjl, {\bf 334}, L103
%
\bibitem[Brand \etal{} 1989]{brand89}
Brand, P.~W.~J.~L., Toner, M.~P., Geballe, T.~R., Webster, A.~S.,
   Williams, P.~M., \& Burton, M.~G. 1989, \mnras, {\bf 236}, 929
%
\bibitem[Burton \& Allen 1994]{burton94}
Burton, M.~G. \&  Allen, D.~A. 1994, 
   in proc.\ ``Infrared Astronomy with Arrays: The Next Generation'', 
   ed.\ I.~S. McLean, (Kluwer, Dordrecht), p61
%
%
%
\bibitem[Dgani \etal{} 1993]{dgani93} 
Dgani, R., Walder, R., \& Nussbaumer, H. 1993, \aap, 267, 155
%
\bibitem[Dougados \etal{} 1993]{dougados93}
Dougados, C., Lena, P., Ridgway, S.~T., Christou, J.~C.,
   \& Probst, R.~G. 1993, \apj, {\bf 406}, 112
%
\bibitem[Draine \& McKee 1993]{draine93}
Draine, B.~T. \& McKee, C.~F. 1993, \araa, {\bf 31}, 373
%
\bibitem[Edwards \etal{} 1993]{edwards93}
Edwards, S., Ray, T.~P., \& Mundt, R. 1993,
   in {\it Protostars and Planets III}, eds.\ E.~H. Levy 
   \& J.~I. Lunine, (Tucson: Univ.\ of Arizona Press), p567
%
\bibitem[Erickson \etal{} 1982]{erickson82}
Erickson, N.~R., Goldsmith, P.~F., Snell, R.~L.,
   Berson, R.~L., Huguenin, G.~R., Ulich, B.~L., Lada, C.~J.
   1982, \apj, {\bf 261}, L103
%
\bibitem[Everett \etal{} 1995]{everett95}
Everett, M.~E., DePoy, D.~L., \& Pogge, R.~W. 1995, \aj, {\bf 110}, 1295
%
\bibitem[Garc\'{\i}a-Segura \etal 1995]{garcia95} 
Garc\'{\i}a-Segura, G., Mac Low, M.-M., \& Langer, N. 1996, \aap, {\bf 305}, 229
%
\bibitem[Geballe \etal{} 1986]{geballe86}
Geballe, T.~R., Persson, S.~E., Simon, T., Lonsdale, C.~J, \& McGregor, P.~J.
   1986, \apj, {\bf 302}, 693
%
\bibitem[Genzel \etal{} 1981]{genzel81}
Genzel, R., Reid, M.~J., Moran, J.~M., \& Downes, D. 1981,
   \apj, {\bf 244}, 884 
%
\bibitem[Genzel \& Stutzki 1989]{genzel89}
Genzel, R. \& Stutzki, J. 1989, \araa, {\bf 27}, 41
%
\bibitem[Gezari \etal{} 1992]{gezari92}
Gezari, D.~Y. 1992, \apj, {\bf 396}, L43
%
\bibitem[Herbst \etal{} 1993]{herbst93}
 Herbst, T.~M., Birk, C., Beckwith, S.~V.~W.,
   Hippler, S., McCaughrean, M.~J., Mannucci, F., \& Wolf, J. 1993,
   Proc.\ SPIE {\bf 1946}, ed.\ A.~M. Fowler, p605
%
%
\bibitem[Hollenbach \& McKee 1989]{hollenbach89}
Hollenbach, D., \& McKee, C. F. 1989, \apj, {\bf 342}, 306
%
\bibitem[Hoare \etal{} 1994]{hoare94}
Hoare, M.~G., Drew, J.~E., Muxlow, T.~B., \& Davis, R.~J. 1994,
     \apj, {\bf 421}, L51
%
\bibitem[Klein \etal{} 1994]{klein94}
Klein, R. I., McKee, C. F., \& Colella, P. 1994, \apj, {\bf 420}, 213
%
\bibitem[K\"onigl \& Ruden 1993]{konigl93} K\"onigl, A. \&
Ruden, S.~P., in {\it Protostars and Planets III}, eds.\ E.~H. Levy \&
J.~I. Lunine, (Tucson: Univ.\ of Arizona Press), p. 641.
%
\bibitem[Lane 1989]{lane89}
Lane, A.~P. 1989, in Low Mass Star Formation and
     Early Stellar Evolution, ed.\ B. Reipurth, (ESO, Garching), p. 331.
%
\bibitem[Langer \etal\ 1994]{langer94}
Langer, N., Hamann, W.-R., Lennon, M., Najarro, F.,
     Pauldrath, A.~W.~A., \& Puls, J. 1994, \aap, {\bf 290}, 819.

\bibitem[Lonsdale \etal{} 1982]{lonsdale82}
Lonsdale, C.~J., Becklin, E.~E., Lee, T.~J., \& Stewart, J.~M.
     1982, \aj, {\bf 87}, 1819 
%
\bibitem[Menten \& Reid 1995]{menten95}
Menten, K.~M. \& Reid, M.~J. 1995, \apjl, {\bf 445}, 157
%
\bibitem[Murata \etal{} 1991]{murata91}
Murata, Y., Kawabe, R., Ishiguro, M., Hasegawa, T., \& Hayashi, M.
   1991, in {\it Fragmentation of molecular clouds and star formation},
   eds.\ E. Falgarone, F. Boulanger, \& G. Duvert, IAU Symposium 147,
   (Dordrecht: Kluwer), p357
%
\bibitem[Murata \etal{} 1992]{murata92}
Murata, Y., Kawabe, R., Ishiguro, M., Morita, K., Hasegawa, T., \& Hayashi, M.
   1992, \pasj, {\bf 44}, 381
%
\bibitem[Neckel \& Staude 1995]{neckel95}
Neckel, Th. \& Staude, H.~J. 1995, \apj, {\bf 448}, 832
%
\bibitem[Parmar \etal{} 1994]{parmar94}
Parmar, P.~S., Lacy, J.~H., \& Achtermann, J.~M. 1994, 
   \apj, {\bf 430}, 786
%
\bibitem[Plambeck \etal{} 1982]{plambeck82}
Plambeck, R. L., Wright, M. C. H., Welch, W. J., Bieging,
   J. M., Baud, B., Ho, P. T. P., \& Vogel, S. N. 1982, \apj, {\bf 259}, 617
%
%
\bibitem[Schild \etal{} 1995]{schild95}
Schild, H., Miller, S., \& Tennyson, J. 1996, \aap, in press
%
\bibitem[Schulz \etal{} 1995]{schulz95}
Schulz, A., Henkel, C., Beckmann, U., Kasemann, C., 
   Schneider, G., Nyman, L.~\AA., Persson, G., Gunnarsson, L.~G., \&
   Delgado, G. 1995, \aap, {\bf 295}, 183
%
\bibitem[Shepherd \& Churchwell 1996]{shepherd96}
Shepherd, D.~S. \& Churchwell, E. 1996, \apj, {\bf 457}, 267
%
\bibitem[Snell \etal{} 1984]{snell84}
Snell, R. L., Scoville, N. Z., Sanders, D. B., \& Erickson,
   N. R. 1984, \apj, {\bf 284}, 176
%
\bibitem[Staude \& Els\"asser 1993]{staude93}
Staude, H.~J. \& Els\"asser, H. 1993, {\it Astron.\ Astrophys.\ Rev.\/},
   {\bf 5}, 165
%
\bibitem[Stone \etal{} 1995]{stone95}
Stone, J.~M., Xu, J., \& Mundy, L.~G. 1995, \nature, 
   {\bf 377}, 315 
%
\bibitem[Sugai \etal{} 1994]{sugai94}
Sugai, H., Usuda, T., Kataza, H., Tanaka, M., Inoue, M.,
   Kawabata, H., Takami, H., Aoki, T., \& Hiromoto, N. 1994,
   \apj, {\bf 420}, 746
%
\bibitem[Sugai \etal{} 1995]{sugai95}
Sugai, H., Kawabata, H., Usuda, T., Inoue, M.~Y.,
   Kataza, H., \& Tanaka, M. 1995, \apj, {\bf 442}, 674
%
\bibitem[Taylor \etal{} 1984]{taylor84}
Taylor, K.~N.~R., Storey, J.~W.~V., Sandell, G., Williams, P.~M.,
   \& Zealey, W.~J. 1984, \nature, {\bf 311}, 236
%
\bibitem[Vishniac 1994]{vishniac94}
Vishniac, E. T. 1994, \apj, {\bf 428}, 186
%
\bibitem[Vogel \etal{} 1984]{vogel84}
Vogel, S.~N., Wright, M.~C.~H., Plambeck, R.~L., \& Welch, W.~J 1984,
   \apj, {\bf 282}, 685
%
\bibitem[Wright \etal{} 1983]{wright83}
Wright, M.~C.~H., Plambeck, R.~L., Vogel, S.~N., Ho, P.~T.~P., \&
   Welch, W.~J 1983, \apj, {\bf 267}, L115
%
\bibitem[Wynn-Williams \etal{} 1984]{wynnwilliams84}
Wynn-Williams, C.~G., Genzel, R., Becklin, E.~E., \&
   Downes, D. 1984, \apj, {\bf 281}, 172
%
\end{thebibliography}
\end{document}